\documentclass[11pt]{article}

\usepackage[margin=1in]{geometry} 
\usepackage{amsmath,amsthm,amssymb,amsfonts}
\usepackage{scalerel}
\usepackage{graphicx}
\usepackage{braket}
\usepackage{authblk} %comment out if you do not have the package

\usepackage{url,soul} % not crucial - just used below for the URL

\usepackage{bbold}		% for identity matrix

\usepackage[colorlinks,citecolor=blue,
            backref=section]{hyperref}
\usepackage{hyperref}
\usepackage{cleveref}% http://ctan.org/pkg/cleveref

\usepackage{algorithm,algorithmic}	
	
\theoremstyle{definition}

\newtheorem{corollary}{Corollary}

\newtheorem{definition}{Definition}
\newtheorem{theorem}{Theorem}
\newtheorem{lemma}{Lemma}
\newtheorem{remark}{Remark}

\begin{document}

\title{Quantum state tomography via non-convex Riemannian gradient descent}			
%\title{Theoretical guarantee of quick convergence and error bound for recovering density matrix under noisy data via non-convex Riemannian gradient descent approach}
\author[1]{Ming-Chien Hsu}
\author[1,2]{En-Jui Kuo}
\author[3]{Wei-Hsuan Yu}
\author[4]{Jian-Feng Cai}
\author[1]{Min-Hsiu Hsieh}

\affil[1]{Hon Hai Quantum Computing Research Center, Taipei, Taiwan}
\affil[2]{Joint Center for Quantum Information and Computer Science, NIST and University of Maryland, College Park,
Maryland, USA}
\affil[3]{Department of Mathematics, National Central University, Taiwan}
\affil[4]{Department of Mathematics, Hong Kong University of Science and Technology, Hong Kong}

\date{}

\maketitle
		
\begin{abstract}
The recovery of an unknown density matrix of large size requires huge computational resources. The recent Factored Gradient Descent (FGD) algorithm and its variants achieved state-of-the-art performance since they could mitigate the dimensionality barrier by utilizing some of the underlying structures of the density matrix. 
Despite their theoretical guarantee of a linear convergence rate, the convergence in practical scenarios is still slow because the contracting factor of the FGD algorithms depends on the condition number $\kappa$ of the ground truth state.
Consequently, the total number of iterations can be as large as $O(\sqrt{\kappa}\ln(\frac{1}{\varepsilon}))$ to achieve the estimation error $\varepsilon$.
In this work, we derive a quantum state tomography scheme that improves the dependence on $\kappa$ to the logarithmic scale; namely, our algorithm could achieve the approximation error $\varepsilon$ in $O(\ln(\frac{1}{\kappa\varepsilon}))$ steps. The improvement comes from the application of the non-convex Riemannian gradient descent (RGD). The contracting factor in our approach is thus a universal constant that is independent of the given state. 
%In this work, we employ the non-convex Riemannian gradient descent (RGD) to derive a quantum state tomography scheme that improves the dependence on $\kappa$ to logarithmic scales. 
%In contrast, the contracting factor in our approach is a universal constant so that the rate of convergence in each iteration is robust and independent of the given state. 
%Under quantified conditions, the matrix reconstruction is thus guaranteed to converge in 
%$O(\ln(\frac{1}{\kappa\varepsilon}))$ steps, leading to high efficiency of the algorithm.
Our theoretical results of extremely fast convergence and nearly optimal error bounds are corroborated by numerical results. 

%This barrier can be mitigated by utilizing some of the underlying structures of the density matrix.
%The recent Factored Gradient Descent (FGD) algorithm and its variants show faster estimation than state-of-the-art convex and non-convex algorithms, since FGD has encoded the rank structure within.
%The recent Factored Gradient Descent (FGD) algorithm and its variants achieved state-of-the-art performance since FGD has encoded the rank structure within.
%Despite their success in obtaining a linear convergence rate, the FGD algorithms suffer from the contracting factor in each iteration usually being close to one, which makes the convergence still slow. 
%\st{Situation is worse if the last singular value of the underlying matrix becomes smaller or the condition number becomes large.}
%In this work, we show much improved rate in convergence by having a universal multiplicative contracting factor via a non-convex Riemannian gradient descent (RGD) algorithm.
%The final reconstruction error is theoretically proved to be bounded in terms of the initial input data noise. 
%This bound is at the same order of those theories obtained from other convex optimization approach, and hence is nearly optimal.
\end{abstract}
		
\section{Introduction}
The density matrix is crucial in describing the quantum state in quantum systems.
Knowing the exact form of a density matrix $\rho$ plays an important role in inferring further properties of the system. 
In some cases, depending on the purpose, only the expectation values of some observables are of concern. In such cases, shadow tomography is used with the focus only on predicting some aspects or properties of the density matrices, rather than the whole \cite{aaronson2018shadow, huang2020predicting_shadow, shadow_Nguyen2022_GM}.
However, arguably, it is always desirable to be able to reconstruct the whole density matrix, whether for the sake of comparison or for more general purposes.

Quantum state tomography involves recovering the density matrix from a given collection of measurements \cite{QST2005photonic, teo2013infoIncompleteQST, czerwinski2021QST_POVM_timeDomain}. 
This can be translated into the optimization problem of finding the best solution with information of given input and certain constraints.
The tomography problem can be formulated and solved in different ways, depending on the different aspects, by using the maximum likelihood estimator (MLE) \cite{PRA99_MLE_DM, MLE_PRA2001_64.024102, ML_Qestimation2001, ML_diluted_PRA2007, MLE_glancy2012gradient}, maximal entropy method \cite{MaxEntro_PRA2013, MaxEntro_PRXqm2021}, and so on.
It has also been shown that the MLE can be converted into a least square minimizer \cite{ML_LS_PRL2012_GaussianN}. 

Most density matrices of interest of size $d\times d$ have some underlying structures.
Such structures, for example, the low rank $r$ structure \cite{flammia2005minimal_M_pure, Gross_2010, Gross_2011, PRA2016_completeM_lowRnk,PRL2016_FisherInfo_PureSt, wang2018pure_InfoComplete}
or the permutation property \cite{permutation_QST_PRL2010, moroder2012permutation_QST}, can be utilized for efficient matrix construction.
Specifically, the low rank $r$ matrices are suitable for compressed sensing frameworks, given that the number of Pauli observable measurements, $m\sim O(rd)$ (ignoring some $\log d$ dependence), is sufficient to recover the density matrix $\rho$, instead of having to find the full $d^2$ information set \cite{Gross_2010, Gross_2011, PRL12_108.170403_Exp_QST_CS, PRA13_87.030102_QST_CM_CS, kalev2015QST_positivity_CS, steffens2017exp_QST_CS}. 
Guaranteed reconstruction is reliant on the restricted isometry property (RIP), which is proven to exist for ordinary Pauli observable measurements \cite{liu2011universal}. 

Since the dimension of the matrix $d=2^k$ grows exponentially with the qubit number $k$, the complexity of the reconstruction increases very quickly.
The fact that, up to now, experimental demonstrations of tomography have only been performed for small qubit numbers \cite{exp_QST_PRL2014, exp_QST_nc2017}, shows the difficulties.
When the matrix dimension $d$ is large, two aspects are particularly relevant in deciding the quality of the tomography. 
One is the sample complexity and the other is the time complexity.
In addition, the algorithm used to recover the density matrix must also guarantee the accuracy.

Sample complexity relates to the fundamental question, how many copies of $\rho$ are necessary and sufficient to determine a state \cite{SampleQST2015arxiv_Haah}. 
Theoretically, according to the positive operator-value measurement (POVM) scheme, Ref.\ \cite{SampleQST2015arxiv_Haah} showed that $O(dr\log(d/\epsilon)/\epsilon)$ copies of $\rho$ are sufficient for tomography to obtain a description with $1-\epsilon$ fidelity, and the necessary lower bound is $\Omega(\frac{rd}{\epsilon\log(d/r\epsilon)})$. In another study,
\cite{Sample_yuen2022improved} improved the lower bound by changing the number of copies to $\Omega(rd/\epsilon)$.
For Pauli measurement, \cite{Sample_yu2020_Pauli}
showed that $O(\frac{10^k}{\delta^2})$ copies are sufficient to
can accomplish the tomography of a $k$-qubit system down to a trace distance error $\delta$.

Time complexity determines the efficiency of an algorithm, which is crucial for matrix recovery in practical applications.
The computational time can be slow for
calculations involving the entire matrix, especially when the system size is large.
Many standard and state-of-the-art algorithms require solving the eigen systems or doing the singular value decomposition (SVD), especially when they involve projection related to the eigenspectrum 
\cite{ML_diluted_PRA2007, OMS2016_PGD_DS, PGD2017_bolduc_QST}, singular value contracting operator \cite{zheng2016r_CS_QST_Measure_sets, zhang2017efficient_HD_ADMM, hu2019reconstructing_14qubit_CS_ADMM}, 
or a unitary transformation of eigenbasis \cite{DIA_PRA2001, ML_LS_PRL2012_GaussianN}.
Both SVD or eigenvalue decomposition have time complexity $O(d^3)$ and thus can be slow.
Other time consuming operations involving the full $d\times d$ matrix include Hessian calculation \cite{GD_PRA2017_superfast}
and matrix inverse \cite{qi2013QST_LRE, li2014robust_ADMM_Pinvserse, li2016improved_ADMM_QST}.
Although the efficiency can be improved in 
\cite{GD_PRA2017_superfast} by switching over from initial costly rapid descent and computing less costly proxy for Hessian \cite{GD_PRA2017_superfast}, it is still heuristic and provides no theoretical guarantee of performance and convergence yet.
Some extension of the matrix inversion case \cite{qi2013QST_LRE} can also improve the efficiency \cite{hou2016full_14qubit_GPU}. 
However, this relies on the graphical-processing unit (GPU) and the linear matrix inversion of the full matrix is not an efficient approach from the algorithmic point of view.
Without utilization of the structures behind the matrix, these algorithms tend to be slow in recovering matrices when the system is large.

A good algorithm should guarantee both the accuracy and efficiency in finding the answer. 
The difference between the final constructed matrix $\hat{\rho}$ and the underlying true density matrix $\rho$ ultimately contains both the error due to the algorithm itself and the error intrinsic to the input measurement data.
The analysis and control of the error bound of this estimated difference is important for the correctness of the algorithm. 
The metric of the error can vary from algorithm to algorithm.
The error bound is shown in nuclear norm in the projected least squares error approach \cite{Tropp_2020fastLS}.
In the convex optimization approach within the compressed sensing framework, the error bounds are shown in both the nuclear norm \cite{candes2011tight,Flammia_2012} and the Frobenius norm \cite{candes2011tight}.

Since the difficulty in tomography is largely caused by the limitations of the algorithms, it is important to find more efficient algorithms. 
Time complexity can be reduced if the underlying structure of the matrices can be utilized. 
Since the density matrices of interest are mostly of low rank, non-convex approaches having the rank structure inherent to the algorithm can perform much better \cite{zhao2015nonconvex, DroppingConvex2016, sun2016guaranteed_NC, tu2016low_Procrustes, ge2017_NC_unified,  park2018finding_NCMF_provably}.
In particular, \cite{kyrillidis2018provable} adopted the non-convex projected Factored Gradient Descent (FGD) to do the tomography.
The Momentum-Inspired version (MIFGD) \cite{kim2021fast} and the stochastic version \cite{kim2022local_stochastic} are the further improved variants of the FGD.
Their results indeed confirm that the FGD method outperforms other approaches,  especially when there is an increase in system size.
This process, however, ignores the eigenvalue dependence during factorization; therefore, each update is heavily dependent on the condition number of the underlying matrix. Moreover, the minimization of errors in each step is related to the eigenvalues and the contracting factor is close to 1. Therefore, it still takes numerous iterations to obtain the final estimation. 

In this paper, we use a much more efficient non-convex Riemannian Gradient Descent (RGD) algorithm that can overcome these difficulties, while still maintaining high guaranteed accuracy.
The RGD algorithm has proven to be both useful and efficient in both matrix recovery problems \cite{Cai15RG} and matrix completion problems \cite{Cai16RG_completion}. 
Its success comes from suitably taking care of the eigenvalues (or singular values in general) in each iteration, so that much more efficient convergence can be expected, while maintaining high accuracy. 
The results show that it takes logarithmic steps to achieve the desired accuracy and that nearly optimal error bounds under noise are guaranteed.

The rest of the paper is structured as follows. In Sec.\ \ref{S:overview} we give an overview of the main results and in Sec.\ \ref{S:technical} we discuss the technical contribution. The related work is reviewed in Sec.\  \ref{S:related}.
A preliminary background is presented in Sec.\ \ref{S:preliminary}. The RGD algorithm and the main results are illustrated in Sec.\ \ref{S:RGD}, while the numerical results are shown in Sec.\ \ref{S:numerical}. Finally, some conclusions are offered in Sec.\ \ref{S:conclusion}.

\subsection{Overview of the main results\label{S:overview}}
The aim of quantum tomography is to recover an unknown density matrix $\rho$ of size $d\times d$ from the measurement outcome $y\in\mathbb{R}^m$, where 
%corresponding to the Pauli measurement on $\rho$. 
the $i$-th component $y_i$ of $y$ corresponds to the expectation value $\operatorname{Tr}(S_i \rho)$ of one sampled Pauli observable $S_i$.
Since most, if not all, of the density matrices of interests are of low rank, we assume $\rho$ to be of rank $r$.
Let $\mathcal{A}$ denote the Pauli sampling which is the mapping acting on $\rho$ to get $m$ collections of the expectation values of Pauli observables. 
Since the measurements inevitably carry noise $z$, the measurement result is written as $y = \mathcal{A}(\rho)+z \in \mathbb{R}^m$.
With $y$ as the input, the quantum state tomography problem could be reformulated and relaxed as an optimization; namely, minimizing the function 
$f(X):=\frac{1}{2}\|y-\mathcal{A}(X)\|_2^2$ over all matrices $X$ such that 
$\operatorname{rank}(X)\le r$. 

This is a non-convex problem that can be efficiently solved with the RGD algorithm. 
The initial guess $X_0$ is chosen to be the rank $r$ approximation of $\mathcal{A}^\dagger(y)$ from the measurement vector $y$. 
Suppose that the noise $z\in\mathbb{R}^m$ obeys the condition $\|\mathcal{A}^\dagger(z)\| \le \lambda$.
The power of the RGD algorithm is shown by the error analysis and time complexity analysis in the following theorem and corollary.

\begin{theorem}
  \label{Thm:rnk_r(simplified)}
  (Simplified) When provided with a small enough $\lambda$ and a large enough samples $m$, the iterate $X_k$ after $k$ steps of the RGD algorithm is guaranteed to be close to $\rho$ in the Frobenius norm
  \[ \|X_{k} - \rho\|_F \le \|X_0 - \rho\|_F  \cdot 
        \bar{\gamma}^k + C \sqrt{r}\lambda,
  \]
with a universal contracting factor $\bar{\gamma}<1$, which is independent of rank $r$, condition number $\kappa$, and so on, where $C=O(1)$ is constant.
\end{theorem}

Here, $\sigma_1$ and $\sigma_r$ denote the largest and smallest singular values,  respectively. More precisely, the  sufficient conditions for the above convergent guarantee $\bar{\gamma}<1$ are that the sampled number of Pauli observables $m\gtrsim O(\kappa^2 r^2 d \log^6 d)$ and 
each Pauli measurement requires a statistical average from $l\sim d/(r\sigma_1^2 \log^5 d)$ number of  measurements.
With the universal contracting factor $\bar{\gamma}<1$, the time complexity analysis for the superfast convergence is as follows.
\begin{corollary}
\label{Cor:TA_simplified}
(Simplified)
The RGD algorithm outputs the estimated matrix $\hat{\rho}$ with error bound $\| \hat{\rho}-\rho\|_F \le C_1 \sqrt{r}\lambda$   
after
\[ \frac{C_2}{\ln(1/\bar{\gamma})} 
     \ln\left( \frac{ \|\rho\|_F}{r\kappa\lambda} \right)
\]
iteration steps for some positive constants $C_1, C_2$ both being $O(1)$.
\end{corollary}

\subsection{Technical Contribution\label{S:technical}}
Under the conditions of small noise $\lambda\le C_1\sigma_r/\sqrt{r}$ and large enough samples $m\ge C_2\kappa^2 r^2 d\log^6 d$,
the RGD estimated matrix will be close to the underlying density matrix, with a nearly optimal error distance.
The convergence is extremely fast, since the required steps are logarithmic with respect to the final errors. 
Further explanations, as well as some advantages over the non-convex FGD-type algorithms, follow.

\begin{itemize}
\item From the convergent time aspect, the error is reduced by a multiplicative contracting factor in each update, leading to a favorable linear convergence rate.
Specifically, the contracting factor $\bar{\gamma}$ in our RGD algorithm
    %to minimize the error $\| \hat{\rho} - \rho \|_F$ iteratively 
is a universal constant, which is independent of all parameters,  including the RIP constant, the condition number $\kappa$ of the underlying density matrix and so on. 
In other words, the error decays exponentially with a constant factor. 
Together with the fact that we could assure the initial approximation error to be inversely proportional to the condition number $\kappa$, the required total number of iteration steps to achieve the final error $\varepsilon$ 
    is at the order of $O(\ln(\frac{1}{\kappa\varepsilon}))$.
%Since the initial guess estimation error is inversely proportional to the condition number $\kappa$,    the required total number of iteration steps to achieve the final error $\varepsilon$ is at the order of $O(\ln(\frac{1}{\kappa\varepsilon}))$.
%Therefore, the RGD algorithm works nicely for  mixed states and/or  matrices, which is is a big advantage compared to other algorithms.

This logarithmic dependence on $\kappa$ in convergent steps is an exponential improvement over the FGD algorithm and its variants.
Although the FGD-type algorithms can also achieve a linear convergence rate, their iterative contracting factor is not universal. Its form 
can be written as $1-\frac{1}{\kappa^\alpha}$, where $\alpha$ can be improved to 0.5 in some variants. 
This gives a total number of iteration steps 
$O(\kappa^\alpha \ln(1/\varepsilon))$
to achieve the final error $\varepsilon$.

\item In each iteration, the step size of each iteration is determined from an exact line search, since the RGD directly minimizes the object function that is quadratic over the set of matrices. The RGD algorithm is therefore easy to execute, as well as implement. In contrast, each matrix $X$ is factored as the form of $AA^\dagger$ in the FGD-type algorithms such that the objective function is quartic in the factored matrix $A$. This makes it impossible to do an exact line search and therefore some prior knowledge or parameters are required to decide the step size in FGD.  

\item In terms of the estimation error, the recovered matrix is nearly optimal in distance to the underlying ground truth density matrix.
The distance error bound $\varepsilon$ is provided in Frobenius norm, which is tighter than the commonly seen nuclear norm.
The final achievable error bound depends on the initial input noise $z\in\mathbb{R}^m$.
    In the noiseless case, where $z=0$, the error can be reduced to nearly zero with arbitrary precision.
    In the noisy case, the final error bound is at the same order of those best known theoretical results from convex optimization approaches \cite{candes2011tight, Flammia_2012}, and hence are nearly optimal.
    
\end{itemize}

\subsection{Related work\label{S:related}}
Since the work of \cite{Gross_2010, Gross_2011}, the process of convex optimization has been shown to be useful for recovering density matrices, particularly in compressed sensing frameworks. For the noiseless case, $m=cdr\log^2 d$ randomly chosen Pauli expectations can uniquely reconstruct the density matrix with high probability \cite{Gross_2010}. For the noisy case, both
the Dantzig selector and the Lasso have been shown to produce similar error bound results \cite{Flammia_2012}.
Suppose that the true underlying matrix $\rho$ is of rank $r$.
The estimated matrix is denoted through the algorithms by $\hat{\rho}$.
Then provided $m\ge C\frac{1}{\delta^2} r d\log^6 d$ and $\|\mathcal{A}^\dagger(z)\|\le \lambda$, there is a high probability that the error bound in the nuclear norm is $\|\hat{\rho}-\rho\|_{*} \le C r \lambda$. 
In comparison, our results show the error bound in both the Frobenius norm $\|\hat{\rho}-\rho\|_F \le C\sqrt{r}\lambda$ and the nuclear norm $\|\hat{\rho} - \rho\|_* \le C r \lambda$.

In terms of the total sample size needed to achieve the nuclear norm error bound $\varepsilon$, the convex optimization methods (Dantzig and Lasso) require $O((\frac{rd}{\varepsilon})^2\log d)$ copies \cite{Flammia_2012}. Our scheme requires the same total sample size $ml$ but it applies to both nuclear norm error  bounded by $\varepsilon$ and Frobenius norm error bounded by $\varepsilon/\sqrt{r}$. 
The projected least squares (PLS) approach also requires a similar sample size $O((\frac{rd}{\varepsilon})^2\log d)$ for Pauli measurements to have an  accuracy $\varepsilon$ in the nuclear norm \cite{Tropp_2020fastLS}. 
The demonstrated PLS is based on using all Pauli observables, while our compressed sensing method allows more delicate separate treatment for the number of sampled Pauli matrices $m$ and the number of measurements $l$ required for each Pauli observable.

Note that the convex optimization method searches for the solution over $d\times d$ matrices while the RGD algorithm searches for the candidate over a tangent space whose size is $d\times r$. 
In addition, the PLS requires a full matrix SVD whose complexity is $O(d^3)$, while the RGD has a complexity $O(d r^2 + r^3)$ for QR decompositions and a SVD.
This means that the RGD algorithm is much less costly in each iterative step than either the convex optimization or the PLS approach. 
Besides, the logarithmic steps of the RGD make its overall computational demand much less than the other approaches while at the same time obtaining the same order of optimal error bounds.

Non-convex approaches utilizing the low rank structures have also been adopted for tomography in past studies \cite{kyrillidis2018provable, kim2021fast, kim2022local_stochastic}. 
Unlike its convex optimization counterpart, the non-convex approach is usually more efficient in terms of computational resources due to the low rank structure utilized.
In particular, the projected FGD approach and its variants decompose each low rank $r$ density matrix as $\rho=AA^\dagger$ for $A\in \mathbb{C}^{d\times r}$ to maintain low rank structures.
Indeed, faster estimation of quantum states 
is achieved by the variant MiFGD \cite{kim2021fast}, 
compared to state-of-the-art convex \cite{yur2015universal_PD_cvx, diamond2016cvxpy, agrawal2018rewriting_CVX} and non-convex \cite{hazan2008sparseApproxSDP} algorithms, including recent deep learning approaches \cite{gao2017efficient_MB_NN, torlai2018neural_QST, QuCumber_WF_NN19, torlai2020MLQST_NISQ}.
However, the FGD-type algorithms still have some shortages due to the factorization. 
This was shown in the previous section in parallel of our technical contribution.

\section{Preliminary\label{S:preliminary}}
In this section, we first introduce some necessary notations for our problem setting. Then we describe the compressed sensing and the concept of restricted isometry property (RIP). The RIP condition is crucial for the matrix recovery problem. Finally we describe the noise and how to obtain its bound by matrix concentration.

\subsection{Notations\label{s:notation}}
For a matrix $M\in\mathbb{C}^{d\times d}$, its adjoint is denoted as $M^\dagger$. Matrix identity is written as $\mathbf{I}$.
In the quantum system, mostly we discuss the matrices and their mapping in the Hermitian space $\mathbb{H}_d(\mathbb{C}) = \{ M\ |\ M\in \mathbb{C}^{d\times d},\ M = M^\dagger\}$. 
We equip the matrix space with the Hilbert-Schmidt inner product by $\operatorname{Tr}(A^\dagger B)$ between matrices $A$ and $B$.
The Frobenius norm for matrix $M$ is defined by $\|M\|_F := \sqrt{\operatorname{Tr}( M^\dagger M)}$, while the nuclear (or trace) norm is the sum of singular values written as $\|M\|_*$. The spectral norm denoted by $\|M\|$ is the largest singular value of $M$. The maps (or superoperators) acting on matrices are written in the calligraphic font, such as $\mathcal{A}$ representing the linear map $\mathbb{C}^{d\times d}\rightarrow \mathbb{R}^m$ and 
the $\mathcal{I}$ standing for the superoperator identity mapping between matrices.
	
%\section{Measurements from sampling Pauli observables \label{s:Pauli_measure}}

We consider the $k$ qubit system, meaning that the matrix dimension $d=2^k$.
The basic Pauli matrices $\{\sigma_i: i=0, 1, 2, 3\}$ (for single quibt) are defined as 
\[  \sigma_0 = \left( \begin{array}{cc}  
                1 & 0 \\ 0 & 1  
                \end{array}  \right), 
    \sigma_1 = \left( \begin{array}{cc}  
                0 & 1 \\ 1 & 0  
                \end{array}  \right), 
    \sigma_2 = \left( \begin{array}{cc}  0 & -i \\ i & 0  
                \end{array}  \right), 
    \sigma_3 = \left( \begin{array}{cc}  1 & 0 \\ 0 & -1  
                \end{array}  \right).
\]
For $k$ qubit systems, we can construct matrices of the tenor product form $P_1 \otimes P_2 \otimes \cdots \otimes P_k$, where $\otimes$ means tensor product and each $P_i$ is a 2 by 2 matrices chosen from $\{\sigma_0, \sigma_1, \sigma_2, \sigma_3\}$. Each such constructed matrix is called a Pauli observable (or matrix) $W_i$ (where $i\in[d^2]$) and there are $d^2 = 4^k$ total of them.
Pauli matrices are Hermitian $W_i = W_i^\dagger$, and obey the orthogonality relation:
\[  \operatorname{Tr}(W_i W_j) = d\delta_{ij},
\]
where $\delta_{ij}$ is the Dirac delta function.
Therefore, they can form the basis of $\mathbb{H}_d(\mathbb{C})$ and every matrix $X\in\mathbb{H}_d(\mathbb{C})$ 
can be expanded by the Pauli matrices as follows: 
\[  X    
    = \frac{1}{d} \sum_{i=1}^{d^2} W_i \operatorname{Tr}(W_i X) 
    = \sum_{i=1}^{d^2} w_i \operatorname{Tr}(w_i X),
\]
where $w_i =W_i/\sqrt{d}$ is the scaled Pauli matrix and $\operatorname{Tr}(W_i X)/d$ is the coefficient corresponding to each $W_i$ in the expansion. Any density matrix $\rho\in\mathbb{H}_d(\mathbb{C})$ and therefore can be expanded in the same way.

\subsection{Compressed sensing and restricted isometry property}
It is natural to reconstruct the matrix $\rho$
from collected coefficients $\operatorname{Tr}(W_i \rho)/d$ for each $W_i$.
For the interest of quantum system, we consider the case that $\rho$ has the rank $r$. Then $\rho$ only has $(2d-r)r$ degrees of freedom. Therefore, we do not need all the coefficients, i.e. we only need $m = O(rd) \ll d^2$ coefficients corresponding to their Pauli matrices and the reconstruction of the matrix  leads to a compressed sensing problem.    
%, which is the case of interests in most quantum systems, not all coefficients are needed to build the matrix, since in this case $\rho$ only has $(2d-r)r$ degrees of freedom.
%Thus we only need $m = O(rd) \ll d^2$ coefficients corresponding to their Pauli matrices, leading to a compressed sensing reconstruction of matrix. 
We choose $m$ basis elements $\{S_1, S_2, \cdots, S_m\}$ i.i.d. uniformly at random from the Pauli basis set $\{W_1, W_2, \cdots, W_{d^2}\}$.
In the chosen set $\{S_i,\ i\in[m]\}$, we define a linear (sensing) map $\mathcal{A}:\mathbb{H}_d(\mathbb{C}) \rightarrow \mathbb{R}^m$ with its $i$-th component corresponding to $S_i$ as 
\begin{equation}
(\mathcal{A}(X))_i = \sqrt{\frac{d}{m}} \operatorname{Tr}(S_i X),
\label{Eq:A}    
\end{equation}
for $X\in\mathbb{H}_d(\mathbb{C})$.
The outcome $\mathcal{A}(X)\in \mathbb{R}^m$ is a vector of dimension $m$. The corresponding self adjoint operator $\mathcal{A}^\dagger: \mathbb{R}^m \rightarrow \mathbb{H}_d(\mathbb{C})$ is 
\begin{equation}
    \mathcal{A}^\dagger(y) = \sqrt{\frac{d}{m}} \sum_{i=1}^{m} y_i S_i,
    \label{Eq:Adagger}
\end{equation}  
where $y\in\mathbb{R}^m$. 
Therefore, we know that $\mathcal{A}^\dagger \mathcal{A}(X) = \frac{d}{m}\sum_{i=1}^{m} \operatorname{Tr}(S_i X) S_i$. 
Since each $S_i$ is chosen from $\{W_1, W_2, \cdots, W_{d^2}\}$ i.i.d. uniformly at random, the expectation is then  
\[ \mathbb{E}[\mathcal{A}^\dagger 
    \mathcal{A}(X)]
  = \frac{1}{d}\sum_{i=1}^{d^2} 
    \operatorname{Tr}(W_i X) W_i
  = X,
\]
giving back the original $X\in\mathbb{H}_d(\mathbb{C})$.

Once $\{S_i,\  i\in[m]\}$ are sampled, the operator $\mathcal{A}$ and $\mathcal{A}^\dagger$ are fixed.
We can then give an estimate $\hat{\rho}$ of the density matrix $\rho$ from a vector $y\in \mathbb{R}^m$, where each $y_i$ corresponds to the coefficient $\{\operatorname{Tr}(S_i\rho)\ |\ i\in[m]\}$ with respect to $S_i$. 
Note that $\{y_i, \ i\in[m]\}$ come from measurement outcomes and thus may contain noise which will be discussed in Section \ref{Sec:noise}.

One important feature for the linear map $\mathcal{A}$ to allow for an exact or reasonable matrix reconstruction is the restricted isometric property (RIP) defined as
\begin{definition}
  The operator $\mathcal{A}$ is said to have RIP with restricted isometric constant $\delta_r$ if it has the following property 
  \[  (1-\delta_r) \|X\|_F^2 \le \|\mathcal{A}X\|_2^2 
     \le (1+\delta_r) \|X\|_F^2,
  \]
  for all the matrices $X$ subject to  $\operatorname{rank}(X)\le r$. The RIP has the fact that $\delta_{r'} \le \delta_r$ if $r' \le r$. 
\end{definition}

The RIP in some sense tells that the combined operator $\mathcal{A}^\dagger \mathcal{A}$
is nearly a superoperator identity when acting on matrix of rank at most $r$.
The Pauli observable measurement defined by $\mathcal{A}$ in Eq.\ (\ref{Eq:A})
is guaranteed to have the RIP with high probability provided that $m=O(rd\log^6 d)$, according to the following theorem. 

\begin{theorem} \cite{liu2011universal} \label{Thm: RIP_pauli}
  Fix some constant $0\le \delta < 1$. We iid uniformly sample $\{S_1, \cdots, S_m\}$ from Pauli matrices $\{W_1, \cdots, W_{d^2}\}$ and define the map $\mathcal{A}$ as Eq.\ (\ref{Eq:A}). Let $m = C \cdot rd \log^6 d$ for some constant $C= O(1/\delta^2)$ depending only on $\delta$. Then over the choice of $(S_1, \cdots, S_m)$, the map $\mathcal{A}$ satisfies the RIP with high probability over the set of all $X\in \mathbb{C}^{d\times d}$ such that $\|X\|_* \le \sqrt{r} \|X\|_F$. Furthermore, the failure probability is exponentially small in $\delta^2 C$.
\end{theorem}

Note that the set of matrices with $\|X\|_* \le \sqrt{r} \|X\|_F$ contains all matrices of rank at most $r$.
Due to the guaranteed RIP (with high probability) of the Pauli measurement, matrices of rank $r$ can be recovered via some suitable optimization approach. 

\subsection{Sampling coefficients with noise \label{Sec:noise}}
Due to the probabilistic nature of quantum phenomena, we can only get the coefficients $\operatorname{Tr}(S_i\rho)$ for a density matrix $\rho$ 
from the statistical frequency average of several measurement outcomes.
Let $y\in\mathbb{R}^m$ denote a scaled vector by collecting all the $m$ measurement results corresponding to $\{\operatorname{Tr}(S_i\rho)\ |\ i\in [m]\}$.
Since measurements almost surely introduce errors, we write $y=\mathcal{A}(\rho)+z$, where $\mathcal{A}$ is defined in Eq.\ (\ref{Eq:A}) and $z$ denotes the noise.

In the following lemma, we show that in fact the noise can be bounded when the number of measurements is large enough, according to the  concentration properties of random variables \cite{tropp2012user, Flammia_2012, Tropp_2020fastLS}. 

\begin{lemma}
\label{Lem:Az_bound}
Let $\mathcal{A}$ be the operator defined in Eq.\ (\ref{Eq:A}) and $y=\mathcal{A}(\rho)+z\in\mathbb{R}^m$ be the corresponding Pauli measurement vector for a density matrix $\rho$, where $z$ denotes the noise. 
Let $\lambda<1$ be a constant.
Then provided that  the number of sampled Pauli observables $m= O(\frac{1}{\delta^2}rd\log^6(d))$
and the number of measurements for each Pauli observable 
$l=O(\frac{\delta^2}{r}\frac{d/\lambda^2}{\log^5 d})$
such that $ml=C d(d+1)\log d/\lambda^2 = C^\prime d^2 \log(d)/\lambda^2$, then 
$\|\mathcal{A}^\dagger (z)\| \le \lambda$  is satisfied  
with probability at least $1-d^{1-C}$.
\end{lemma} 

Therefore, as long as the number of total measurements $ml$ is large enough, we can have  the noise $z$ bounded by
$\|\mathcal{A}^\dagger (z)\| \le \lambda$
for a desired bound $\lambda$ with high probability. This can be beneficial in the matrix reconstruction procedure. 
The proof of Lemma \ref{Lem:Az_bound} is shown in the Appendix \ref{Appendix:z_noise}.

\section{Solving Tomography via non-convex Riemannian gradient descent Approach\label{S:RGD}}
For quantum systems, the density matrix has dimension $d\times d$ exponentially large with respect to qubit number $k$.
The matrix estimation or recovery usually requires large resources either by exact recovery or convex optimization.
In tomography problems, the to-be-solved matrix of interests is usually of special structures, such as low rank property.
Here we utilize the low rank $r$ structure and use an efficient non-convex optimization approach called the Riemannian gradient descent (RGD) algorithm to solve the problem. 

The tomography problem here corresponds to estimating a density matrix from a given input vector $y$ related to the underlying density matrix $\rho$.
The matrix $\rho$ is a fixed but unknown density matrix to be determined. 
The vector $y = \mathcal{A}(\rho)+z \in\mathbb{R}^m$ is ideally only coming from the result of the sensing operator $\mathcal{A}$ defined in Eq.\ (\ref{Eq:A}) corresponding to the expectation values of sampled Pauli observables but inevitably it contains the noise $z$. The $i$-th component $y_i$ corresponds to the sampled Pauli observables $S_i$. Once the Pauli matrices $\{S_i\}$ is chosen, the correspondence is fixed and the fixed chosen set $\{S_i\}$ is then used for any optimization approach to perform the matrix recovery.  
The noise $z$ is supposed to be bounded under the mapping $\mathcal{A}^\dagger$ such that spectral norm  $\|\mathcal{A}^\dagger(z)\| \le \lambda$ being bounded by matrix concentration as mentioned above.
With these, the tomography problem of solving the density matrix $\rho$ is formulated and relaxed to the following non-convex optimization problem
\begin{equation}
    \displaystyle \min_{X\in\mathbb{H}_d}\   
        f(X) := \frac{1}{2}||y-\mathcal{A}(X)||_2^2\ \ \text{subject to } 
        \operatorname{rank}(X)\le r,
    \label{Eq.OptProb}
\end{equation}
where both the constraints of unit trace and semidefinite positiveness $X\succeq 0$ are relaxed.
The relaxation of the unit trace constraint is reasonable 
since the Frobenius norm and the nuclear norm distance of the final estimated $\hat{\rho}$ to the underlying $\rho$ is small as well as the fact that the trace of $\hat{\rho}$ can be influenced by the noise $z$ level and can deviate from 1.
The condition $X\succeq 0$ is also relaxed since the eigenvalues and singular values are the same for the underlying to-be-solved $\rho$ and the final estimated matrix will automatically satisfy the semidefinite positiveness.
The noise bound $\lambda$ is also not used as a constraint; in contrast, the condition $\|\mathcal{A}^\dagger(z)\| \le\lambda$ is used to analyze the final corresponding error bound of the result.

In this section, we use the RGD algorithm to solve the optimization problem.
{Let} the output of the optimization be $\hat\rho$, and then the Frobenius norm of $\hat{\rho}$ to the true density $\rho$ will be bounded. It will be proved in the following.

\begin{algorithm}[H]
    \caption{RGD Algorithm solving matrix recovery}
    \label{alg:RGD}
\begin{algorithmic}
\STATE Input: $\mathcal{A}, y$ and rank $r$.
\STATE Initialize $X_0$ and do the singular value decomposition (SVD) $X_0 = U_0 \Sigma_0 V_0^\dagger$.
\FOR{$k=1, \ldots$}
  \STATE 1. find the direction 
            $G_k = \mathcal{A}^\dagger(y - \mathcal{A}(X_k))$ 
  \STATE 2. determine the step size 
            $\alpha_k = \frac{\|\mathcal{P}_{T_k}(G_k)\|_F^2}
  {\|\mathcal{A}\mathcal{P}_{T_k}(G_k)\|_2^2}$.
  \STATE 3. the intermediate matrix on the tangent space 
            $W_k = X_k + \alpha_k \mathcal{P}_{T_k}(G_k)$.
  \STATE 4. update the estimated matrix $X_{k+1} = \mathcal{H}_r(W_k)$ 
\ENDFOR
\STATE Output: $\hat{\rho} = X_k$ after $k$ steps when the stopping criteria is met.
 
\end{algorithmic}
\end{algorithm}

The RGD algorithm is an iterative algorithm to solve the optimization problem (\ref{Eq.OptProb}) with well defined and fixed $\mathcal{A}$ and requires input $y$.
In each iteration step, the estimated matrix is updated via projected gradient descent along the tangent space of the previous step with suitable step size.
Suppose the $k$-th step matrix $X_k$ has singular value decomposition (SVD) $X_k = U_k\Sigma_k U_k^\dagger$, and the projections onto its column and row space are denoted as $P_{U_k} = U_k U_k^\dagger$ and $P_{V_k} = V_k V_k^\dagger$ respectively.
The tangent space $T_k$ at the current step $X_k$ is determined by the spanning of the column and row space as 
\[  T_k = \{X\in\mathbb{H}_d\ |\ (\mathbf{I}-P_{U_k})X(\mathbf{I}-P_{V_k})=0\},
\]
and the corresponding projection $\mathcal{P}_{T_k}$ is  
\[  \mathcal{P}_{T_k}: X \mapsto P_{U_k}X + X P_{V_k} - P_{U_k}X P_{V_k}.
\]
With suitable step size $\alpha_k$, the projected gradient descent along the tangent space gives $W_k$. 
Then we apply the hard thresholding operator $\mathcal{H}_r$ on $W_k$ to get the updated $(k+1)$-th step matrix $X_{k+1}$ which is still of rank $r$.
The operator $\mathcal{H}_r$ acting on any matrix $X$ is to produce its truncated rank $r$ approximation $X_r$ which preserves the top $r$ singular values $\sigma_1, \sigma_2, \cdots \sigma_r$ in decreasing order and the corresponding singular vectors. All the left $\sigma_{r+1}, \cdots$ are discarded under $\mathcal{H}_r$.
Note that matrices at each tangent space $T_k$ have rank at most $2r$, and therefore the computation complexity for $\mathcal{H}_r$ is low for low rank $r$ cases. 

\subsection{The main theorem for recovering the density matrix \label{Sec:mainThm_rank_r}}
In the following main theorem, we consider the underlying density matrix $\rho$ to be of rank $r$ with singular values $\sigma_1, \sigma_2, \cdots, \cdots \sigma_r >0$ in decreasing order.
Since density matrix is positive semidefinite, its eigenvalues are the same as its singular values. Therefore, we also have $\sum_{i=1}^r \sigma_i = 1$ and recovering the SVD of $\rho$ is the same as recovering $\rho$ itself.
Starting from the initial point $X_0 = \mathcal{H}_r(\mathcal{A}(y))$ from the measurement $y$ as input, we obtain the estimated matrix $\hat{\rho}$ via the RGD algorithm.
The estimated $\hat{\rho}$ can be arbitrarily close to the true matrix $\rho$ for noiseless case.
Suppose the noise $z$ satisfies $\|\mathcal{A}^\dagger(z)\|\le\lambda$.
We can upper bound the error of the estimated $\hat{\rho}$ in terms of $\lambda$.
The theorem still holds when the noise bound $\lambda$ becomes 0 (or $z=0$) and goes back to the noiseless case. We defer the proof of Theorem \ref{Thm:rnk_r} to the latter part in Sec.\ \ref{Proof:Thm_rank_r}.

\begin{theorem}
  \label{Thm:rnk_r}
  (main result) Let $\rho$ be a density matrix of rank $r$ with a measurement result  
  $y=\mathcal{A}(\rho)+z \in \mathbb{R}^m$ where the mapping $\mathcal{A}$ is defined in Eq.\ (\ref{Eq:A}), and the noise $z$ is supposed to obey $\|\mathcal{A}^\dagger(z)\|\le\lambda$.
  Denote condition number of $\rho$ be $\kappa := \sigma_1/\sigma_r$, where $\sigma_1$ and $\sigma_r$ denote the first and the $r$-th singular value of $\rho$.
  Then there exists constants $C_1, C_2 >0$ such that when provided $\lambda \le C_1 \sigma_r/\sqrt{r}$ and 
  $m \ge C_2 \kappa^2 r^2 d\log^6 d$, then the $k$-th iterates of the RGD algorithm \ref{alg:RGD} with initial point $X_0 = \mathcal{H}_r(\mathcal{A^\dagger}(y))$
  has rank at most $r$ and is
  guaranteed to be close to the true $\rho$ in Frobenius norm distance bounded as 
  \begin{equation}
     \|X_{k} - \rho\|_F \le \|X_0 - \rho\|_F \cdot \bar{\gamma}^k
     + \frac{2\sqrt{2r}\lambda}{1-\delta_{3r}} \left(\frac{1}{1-\bar{\gamma}}\right),
    \label{Eq:XkFro_rnk_r}    
  \end{equation}
  where the contracting factor $\bar{\gamma} < 1$ is a universal bound in all steps and $\delta_{3r}$ is the RIP constant of $\mathcal{A}$. 
\end{theorem}

%%%  X0 error can be minimized by RGD
\begin{remark}
Some may wonder the initial $X_0$ may be good enough. 
However, we point out that there exists regions that the estimation error $\|X_0 - \rho\|_F$ due to the initial point choice $X_0=\mathcal{H}_r(\mathcal{A}^\dagger(y))$ can be further reduced by the RGD algorithm \ref{alg:RGD}, in particular for small noise cases. To see this, we demonstrate the bound of the errors due to the initialization and the accumulated error of the iterates separately.

Now we show that provided the case $\lambda \le \sigma_r/(40\sqrt{2r})$ and $\delta_{3r}\le 1/(80\sqrt{r}\kappa)$ where $\kappa$ the condition number, consistent with the conditions in Theorem \ref{Thm:rnk_r}, then we have the universal contracting factor upper bound $\bar{\gamma} < 0.23$. Therefore, the last term in Eq.\ (\ref{Eq:XkFro_rnk_r}) being the iterate error is bounded by
  \[ \frac{2\sqrt{2r}\lambda}{1-\delta_{3r}} \left(\frac{1}{1-\bar{\gamma}}\right) \le 3.72 \sqrt{r}\lambda \le 0.0658 \sigma_r,
  \] 
while the initialization error from $X_0=\mathcal{H}_r(\mathcal{A}(y))$ is bounded from Eq.\ (\ref{Eq:initX0}) according to Lemma \ref{Lem:X0_bound} in the appendix such that 
    \[  \|X_0 - \rho\|_F 
        \le 2\delta_{2r}\|\rho\|_F +     
                2\sqrt{2r}\lambda
        \le 2\delta_{2r}\sqrt{r}\sigma_1 +          
            2\sqrt{2r}\lambda
        \le 0.075 \sigma_r.
    \]
In overall, since the contracting error $\bar{\gamma}<1$  
makes $\|X_0 - \rho\|_F \cdot \bar{\gamma}^k$ to vanish, 
the $\|X_{k} - \rho\|_F$ is dominated by the iterate error $\le 0.0658 \sigma_r$ and is smaller than the initialization error $0.075 \sigma_r$ in this case.
Further note that with smaller noise $\lambda$, the RGD algorithm \ref{alg:RGD} can greatly reduce the final error by vanishing the initialization error $\|X_0 - \rho\|_F \cdot \bar{\gamma}^k$ and at the same time maintaining the accumulation error bounded as $C\sqrt{r}\lambda$ where $C=O(1)$
such that $C\sqrt{r}\lambda$ is smaller than the initial $\|X_0 - \rho\|_F$.
\end{remark}

%% -------- condition for samples ----- %
\begin{remark}
Now we discuss the conditions for the RGD to have the universal contraction factor $\bar{\gamma}<1$ and the guaranteed error bounds $\|X_k - \rho\|_F = C\sqrt{r}\lambda$ and $\|X_k -\rho\|_* = C r\lambda$.
The noise bound $\lambda\le C_1\sigma_r/\sqrt{r}$ requires that $ml= C^\prime d^2 \log(d)/\lambda^2 \ge C^{\prime\prime} r d^2 \log(d)/\sigma_r^2$, according to Lemma \ref{Lem:Az_bound}. This is the same order as the total measurement number required both in the convex optimization approach \cite{Flammia_2012} and
in the projected least square approach \cite{Tropp_2020fastLS}.

The other condition about the number of sampled Pauli observables 
$m \ge C_2 \kappa^2 r^2 d \log^6 d$ is equivalent to have the RIP constant 
$\delta_{3r}\le \frac{C^{\prime\prime\prime}}{\kappa \sqrt{r}}$ for the mapping $\mathcal{A}$.
This in turns requires the number of measurement for each Pauli observable to be $l = O(  d/(\kappa^2 r^2 \lambda^2 \log^5 d)) = O(d/(\sigma_1^2 r \log^5 d))$ for the statistical average.

\end{remark}

%%%  convergent rate  (time analysis)
\begin{corollary}
\label{Cor:Crate}
Let $\rho$ be a density matrix of rank $r$ with measurement $y= \mathcal{A}(\rho)+z$ with $\mathcal{A}$ defined in Eq.\ (\ref{Eq:A}) and the noise $z$ satisfying $\| \mathcal{A}(z)\|\le \lambda$. Under the conditions in Theorem \ref{Thm:rnk_r}, there exists positive constants $C_0, C_1, C_2$ all being $O(1)$ and $C_1 < C_2$ such that the RGD algorithm \ref{alg:RGD} can output the estimated density matrix $\hat{\rho}$ close to $\rho$ of rank $r$ obeying
\[  \frac{\| \hat{\rho}-\rho\|_F}{\|\rho\|_F} \le 
    C_2 \frac{\sqrt{r}\lambda}{\|\rho\|_F},
\]
after
\begin{equation}
 \frac{1}{\ln(1/\bar{\gamma})}\left(
          \ln\left( \frac{2 C_0 \|\rho\|_F}{r\kappa\lambda} +
                2\sqrt{2} \right)
        - \ln(C_2 - C_1)  \right)
 \label{Eq:log_k_steps}    
\end{equation}
iteration steps, where $\kappa:=\sigma_1/\sigma_r$ is the condition number of $\rho$, and $\bar{\gamma}$ is a universal constant smaller than 1.

When applied to the noiseless case, that is $\lambda =0$, the RGD algorithm outputs $\hat{\rho}$ with 
\[  \frac{\| \hat{\rho} - \rho\|_F}{\|\rho\|_F}  \le
    \varepsilon,
\]
after $\ln\left( \frac{C_0}{\sqrt{r} \kappa \varepsilon}\right)/\ln(\frac{1}{\bar{\gamma}})$ iteration steps.
\end{corollary}

\begin{proof}
  According to Theorem \ref{Thm:rnk_r}, the RGD algorithm has the iterate $X_k$ after $k$ iteration steps to be bounded from Eq.\ (\ref{Eq:XkFro_rnk_r})
  \[ \|X_{k} - \rho\|_F \le \|X_0 - \rho\|_F \cdot \bar{\gamma}^k
     + \frac{2\sqrt{2r}\lambda}{1-\delta_{3r}} \left(\frac{1}{1-\bar{\gamma}}\right),
  \]
  with initial point bounded by Eq.\ (\ref{Eq:initX0}) according to Lemma \ref{Lem:X0_bound} in the appendix
    \[  \|X_0 - \rho\|_F \le 2\delta_{2r}\|\rho\|_F +     
                2\sqrt{2r}\lambda.
    \]
  The existence of the universal constant $\bar{\gamma} < 1$ is guaranteed when the condition $\delta_{2r}\le\delta_{3r}\le C_0^\prime\frac{1}{\kappa \sqrt{r}}$ is satisfied
  where $0< C^\prime_0 < 1$ is usually small so that $1/(1-\delta_{3r})$ is $O(1)$.
  Besides, since the positive $\bar{\gamma}$ is usually not close to 1, we have the constant $C_1 \ge \frac{1}{1-\bar{\gamma}}\frac{1}{1-\delta_{3r}}$ which is $O(1)$ such that 
  \[  \begin{array}{ccl}
     \frac{\|X_{k} - \rho\|_F}{\|\rho\|_F} 
    &\le& \frac{ \|X_0 - \rho\|_F}{\|\rho\|_F} \cdot \bar{\gamma}^k
    + C_1 \frac{\sqrt{r}\lambda}{\|\rho \|_F} \\

    &\le& \left(2\delta_{2r} + \frac{ 2\sqrt{2r}\lambda}
    {\|\rho\|_F}\right) \cdot \bar{\gamma}^k
     + C_1 \frac{\sqrt{r}\lambda}{\|\rho \|_F}  \\

    &\le& \left(2C^\prime_0\frac{1}{\kappa\sqrt{r}} + \frac{ 2\sqrt{2r}\lambda}
    {\|\rho\|_F}\right) \cdot \bar{\gamma}^k
     + C_1 \frac{\sqrt{r}\lambda}{\|\rho \|_F}  \\

    &=& \left( 
    \left( \frac{C_0\|\rho\|_F}{r \kappa\lambda} + 2\sqrt{2}
    \right) \cdot \bar{\gamma}^k
     + C_1 \right) \frac{\sqrt{r}\lambda}{\|\rho \|_F},
    \end{array}
  \]  
  where $C_0 = 2C^\prime_0$.
  To have $\|X_{k} - \rho\|_F/ \|\rho\|_F \le C_2 \sqrt{r}\lambda/ \|\rho \|_F$, we need
\[   k \ln(\bar{\gamma})
      \le  \ln(C_2 - C_1) - \ln\left( 
     \frac{C_0\|\rho\|_F}{r \kappa\lambda} + 2\sqrt{2} 
    \right).
\]  
  Note that the universal constant $0<\bar{\gamma}<1$ such that $\ln(1/\bar{\gamma})>0$ and thus the iteration steps $k$ as claimed to have the output 
  $\hat{\rho} = X_k$ as desired closeness to $\rho$.
  
  For the noiseless case, $y=\mathcal{A}(\rho)$ equivalent to $\lambda=0$, and thus we have 
  \[ \|X_{k} - \rho\|_F 
        \le \|X_0 - \rho\|_F \cdot \bar{\gamma}^k 
        \le (2\delta_{2r}\|\rho\|_F) \cdot \bar{\gamma}^k,
  \]
  which is equivalent to  
  \[ \frac{\|X_{k} - \rho\|_F}{\|\rho\|_F} 
        \le (2\delta_{2r}) \cdot \bar{\gamma}^k 
        \le (C_0 \frac{1}{\kappa\sqrt{r}}) \cdot \bar{\gamma}^k.
  \]
  Therefore, with 
  $k = \ln\left( \frac{C_0}{\sqrt{r} \kappa \varepsilon}\right)/\ln(\frac{1}{\bar{\gamma}})$ we can have $\|X_{k} - \rho\|_F/ \|\rho\|_F \le \varepsilon$. 
  
\end{proof}

\begin{remark}
    With bounded input noise $\|\mathcal{A}(z)\|_F\le \lambda$, 
    it is theoretically analyzed 
    the sufficient conditions to guarantee the contracting factor and to obtain the error bound.
    In such case, the error in the Frobenius norm is bounded by $\varepsilon = C\sqrt{r}\lambda$ with $C=O(1)$ for rank $r$ density matrices,
    which can be converted via the triangular inequality to become $Cr\lambda$ in the nuclear norm.
    The error bounds here are at the same order of the best bounds from convex optimization in the Frobenius norm \cite{candes2011tight}
    and in the nuclear norm \cite{Flammia_2012}, separately, 
    and hence are nearly optimal.
    In short, the RGD algorithm can determine errors both in the Frobenius norm and the nuclear norm with the same order of optimality as the convex approach, but achieve this result with faster logarithmic steps.
\end{remark}

\begin{remark}
    For comparison, the required number of iterations in FGD has dependence on the condition number $\kappa$ to be $O(\kappa^\alpha)$, where the best $\alpha\sim 0.5$.
    This complexity is much more demanding compared to 
    the $O(\ln(1/\kappa))$ in the RGD. 
    The slow rate in the FGD comes from the fact that the multiplicative factor is actually closer to 1 when the condition number increases. 
    This is because the FGD and its variants absorb the singular values to the factored matrix $A$
    such that the updating rules carry the condition number of matrices and hence are not optimal.
    This results in the FGD type algorithms not working well when the condition number is large.
\end{remark}

\begin{remark}
  In terms of the computational effort, each iteration of the RGD algorithm is also efficient.
Specifically, the RGD algorithm fully exploits the low rank structure and only searches over the low rank tangent spaces. 
The main computation of projection onto the tangent space $\mathcal{P}_{T_k}$
consists of matrix products between matrices of size $d\times d$ and $d\times r$, two QR decompositions of matrices of size $d\times 2r$ and one SVD of a matrix of size $2r\times 2r$; hence the complexity is $O(d^2 r + d r^2 + r^3) = O(d^2 r)$.
In comparison, the expensive SVD for the full matrix of the system size $d\times d$ is not needed in our RGD algorithm.
Therefore, the RGD is much cheaper in terms of computational resources, especially for cases when the rank $r$ is small compared to $d$ which are commonly seen in quantum systems with large qubits.
\end{remark}

\begin{remark}
    Another computational effort that can be saved comes from the much shorter convergent steps in iteration which are $O(\ln(\frac{1}{\kappa\varepsilon}))$ mentioned above.
    Besides, since in all iterative algorithms, it is necessary to calculate the  $\mathcal{A}(M)$ for some matrix $M$ related to $X_k$ each time it is updated, this minimizes the total number of times $\mathcal{A}$ must be calculated which may be relatively expensive in some cases.
\end{remark}

% --------  proof of main theorem (Thm 3)
\subsection{The Proof of Theorem \ref{Thm:rnk_r} \label{Proof:Thm_rank_r}}
\begin{proof}
  To analyze the errors in terms of the initial point and noise, we derive the bound in the following steps.
  \begin{itemize}
      \item We first write write the bound in successive iterative relation, that is
      relating the $(k+1)$-th step bound to the $k$-th step bound. We write the bound $\|X_{k+1} - \rho\|_F\le B_k + B_z$, including two terms. The $B_k$ is written in terms of $\|X_{k} - \rho\|_F$, and has no explicit noise term, so called the noiseless term. The other term $B_z$ is from the noise $z$, so called the noise term.
      
      \item From $X_0=\mathcal{H}_r(\mathcal{A}(y))$ we bound the initial estimation error $\|X_0 - \rho\|_F$. 
      
      \item The next step is to packaging the iterate bound $\|X_{k} - \rho\|_F$ in terms of $\|X_0 - \rho\|_F$. This is to see how the errors propagating along the iteration. Due to this, even each $B_k$ term implicitly includes noise effect.
      
      \item Finally, we find cases with sufficient conditions such that each update has contracting factor $\bar{\gamma}<1$ in the $B_k$ term and equally importantly the accumulated $B_z$ term can be also upper bounded.
  \end{itemize}

  Firstly, to write $\|X_{k+1} - \rho\|_F$ in terms of $\|X_{k} - \rho\|_F$ in each iteration update, we first note that $X_{k+1}=\mathcal{H}_r(W_k)$ is the best rank $r$ approximation for $W_k$, since
  by the Eckart-Young theorem we know that 
  \begin{equation}
     \|X_{k+1} - W_k\|_F \le \| \rho - W_k\|_F.
     \label{Eq:Eckart-rank_r}    
  \end{equation}
  In other words, the Frobenius norm difference of $X_{k+1}$ to $W_k$ is always the smallest among all rank $r$ matrices, including $\rho$ which is rank $r$ by assumption.
  Note that $G_k = -\nabla f(X_k)$ is the gradient descent and $W_k$ is thus the projected gradient descent along the tangent space $T_k$ of the current step $X_k$: 
  \begin{equation}
      \begin{array}{ccl}
      W_k &=& X_k + \alpha_k \mathcal{P}_{T_k}(G_k)  \\
          &=& X_k + \alpha_k \mathcal{P}_{T_k}(\mathcal{A}^\dagger(y -   
                \mathcal{A}(X_k))  \\
          &=& X_k + \alpha_k \mathcal{P}_{T_k}(\mathcal{A}^\dagger\mathcal{A}
          (\rho - X_k) + \mathcal{A}^\dagger(z)).  \\
     \end{array}
     \label{Eq:Wk}
  \end{equation}
  It follows that by triangular inequality
  \begin{equation}
      \begin{array}{cl}
      \|X_{k+1} - \rho\|_F & \le 
            \|X_{k+1} - W_k\|_F + \|W_k - \rho\|_F  
        \le 2 \| \rho - W_k\|_F  \\
       & \le 2 \| (\mathcal{I} - \alpha_k \mathcal{P}_{T_k}\mathcal{A}^\dagger\mathcal{A})(\rho - X_k)\|_F + 
       2 \| \alpha_k \mathcal{P}_{T_k}\mathcal{A}^\dagger(z)\|_F
       =: B_k + B_z, 
      \end{array}
      \label{Eq:X(k+1)}
  \end{equation}
  where $\mathcal{I}$ is the superoperator identity. 
  We denote the first term $B_k := 2 \| (\mathcal{I} - \alpha_k \mathcal{P}_{T_k}\mathcal{A}^\dagger\mathcal{A})(\rho - X_k)\|_F$ referring to the noiseless case, and the second term
  $B_z := 2 \| \alpha_k \mathcal{P}_{T_k}\mathcal{A}^\dagger(z)\|_F$ the effect arising from the noise. 
  
  {\it Bounding the $B_z$ term}: We note that
  $\operatorname{rank}(\mathcal{P}_{T_k}(M))\le 2r$ for any matrix $M\in \mathbb{C}^{d\times d}$ 
  so that 
  \begin{equation}
       \|\mathcal{P}_{T_k}\mathcal{A}^\dagger(z)\|_F
         \le \sqrt{2r}\|\mathcal{P}_{T_k}\mathcal{A}^\dagger(z)\| 
        \le \sqrt{2r}\|\mathcal{A}^\dagger(z)\| \le \sqrt{2r}\lambda,
       \label{Eq:Tk_noiseBound}      
  \end{equation}
  where the last inequality comes the assumption on noise bound.
  Further note that the theorem assumption on $m$ translates into $m=O(\frac{1}{\delta^2} rd \log^6 d)$ such that the mapping $\mathcal{A}$ satisfies the RIP with constant $\delta=O(\frac{1}{\kappa\sqrt{r}})$.   
  Then by applying the RIP, we can have 
  \[  (1-\delta_{2r})\|\mathcal{P}_{T_k}(G_k)\|_F^2 \le \|\mathcal{A}\mathcal{P}_{T_k}(G_k)\|_2^2 
  \le(1+\delta_{2r})\|\mathcal{P}_{T_k}(G_k)\|_F^2.
  \] 
  Therefore, the step size $\alpha_k$ is also bounded as 
  \[ \frac{1}{1+\delta_{2r}} \le 
          \alpha_k = \frac{\|\mathcal{P}_{T_k}(G_k)\|_F^2}
    {\|\mathcal{A}\mathcal{P}_{T_k}(G_k)\|_2^2} \le \frac{1}{1-\delta_{2r}},
  \]
  leading to
  \[ B_z \le 2 |\alpha_k|       
        \|\mathcal{P}_{T_k}\mathcal{A}^\dagger(z)\|_F
            \le \frac{2\sqrt{2r}\lambda}{1-\delta_{2r}}. 
  \]
  
  {\it Bounding the $B_k$ term}: For the noiseless part, we outline and summarize results according to steps in  \cite{Cai15RG}. 
  By triangular inequality, we have 
  \[  \begin{array}{ccl}
  B_k &\le& 2  \| (\mathcal{I} - \mathcal{P}_{T_k})
                 (\rho - X_k)\|_F    
     + 2 \| (\mathcal{P}_{T_k} - \alpha_k \mathcal{P}_{T_k}
        \mathcal{A}^\dagger\mathcal{A}\mathcal{P}_{T_k})
        (\rho - X_k)\|_F  \\
    &+& 2 \| \alpha_k \mathcal{P}_{T_k}\mathcal{A}^\dagger\mathcal{A}
         (\mathcal{I}-\mathcal{P}_{T_k})(\rho - X_k)\|_F \\
    &=:& B_1 + B_2 + B_3.       \\
  \end{array}
  \]
  Summarizing according to \cite{Cai15RG}, we can bound each term as
  \[ \begin{array}{ccl}
     B_1 &\le&  \frac{2}{\sigma_r}\|X_k - \rho\|_F^2, \\
     B_2 &\le&  \frac{4\delta_{2r}}{1-\delta_{2r}}
            \|X_k - \rho\|_F,\ \text{ and } \\
     B_3 &\le&  \frac{2\delta_{3r}}{1-\delta_{2r}}
                    \|X_k - \rho\|_F. \\
      \end{array}
  \]
  Note that as iteration goes on, these term still implicitly carry noise effect which will be analyzed below.
  
  {\it Bounding $\|X_{k+1}-\rho\|_F$}: Summing up, we have the upper bound in the $(k+1)-$th update step
  \begin{equation}
      \|X _{k+1} - \rho\|_F \le B_k + B_z \le
          \left(\frac{4\delta_{2r} + 2\delta_{3r}}{1-\delta_{2r}} 
        + \frac{2}{\sigma_r} \|X_k - \rho\|_F \right) 
            \|X_k - \rho\|_F
        + \frac{2\sqrt{2r}\lambda}{1-\delta_{2r}}.  
    \label{Eq:NewStepBd}
  \end{equation}  
  
  Secondly, we want to bound the initial estimation error $\|X_0 -\rho\|_F$ where $X_0 =\mathcal{H}_r(\mathcal{A}(y))$. According to Lemma \ref{Lem:X0_bound} shown in the Appendix \ref{Appendix:X0_bound}, we have the bound Eq.\ (\ref{Eq:initX0})
  \[ \|X_0 - \rho\|_F \le 2\delta_{2r}\|\rho\|_F 
     + 2\sqrt{2r}\lambda.
  \]
  
  Thirdly, we show how the noise propagate along the iterations. Since each step will accumulate the noise propagating from the very beginning, we now derive the bound in terms of the previous step and hence of $||X_0 - \rho||_F$ plus the noise contribution. For convenience, we define the following terms 
  \[ \theta = \frac{4\delta_{2r} + 2\delta_{3r}}{1-\delta_{2r}},
     \eta = 4\delta_{2r}\sqrt{r} \frac{\sigma_1}{\sigma_r}, 
     \phi = 4\sqrt{2r} \frac{\lambda}{\sigma_r}, \text{ and }
     \omega = 2\sqrt{2r} \frac{\lambda}{1-\delta_{2r}}.
  \]
  We also define the following terms by recursion relation
  \begin{equation}
    \begin{array}{ccl}
        \gamma_0 &=& \theta + \eta + \phi,       \\
        \gamma_1 &=& \theta + (\eta+\phi)\times\gamma_0 
                    + \frac{\phi}{1-\delta_{2r}}\times \mu_1,   \\
        \gamma_2 &=& \theta + (\eta+\phi)\times\gamma_0\gamma_1 
            + \frac{\phi}{1-\delta_{2r}}\times \mu_2,   \\
        \gamma_3 &=& \theta + (\eta+\phi)\times\gamma_0\gamma_1\gamma_2 
            + \frac{\phi}{1-\delta_{2r}}\times \mu_3,   \\
        &\vdots&        \\
        \gamma_k &=& \theta + (\eta+\phi)\times
            \gamma_0\gamma_1\cdots\gamma_{k-1}
            + \frac{\phi}{1-\delta_{2r}}\times \mu_k,
      \end{array}
    \label{Eq:gamma_recursion}      
  \end{equation}  
  where $\mu_1=1$ and $\mu_{k+1}= 1+\gamma_k\mu_k$ so that 
  \begin{equation}
    \mu_2 = 1+\gamma_1,\ \mu_3 = 1+\gamma_2 + \gamma_2\gamma_1,\ \cdots,\
     \mu_{k+1} = 1+ \gamma_k + \gamma_k\gamma_{k-1} + \cdots 
                  + \gamma_k\cdots\gamma_1.
    \label{Eq:mu_recursion}  
  \end{equation} 
  All these terms are functions of $\theta, \eta$ and $\phi$. In terms of these defined terms, we claim that the matrix distance satisfy the following
  \begin{equation}
      \begin{array}{ccl}
        \|X_1 - \rho\|_F &\le& \|X_0 - \rho\|_F \gamma_0  + \omega \mu_1,   \\
        \|X_2 - \rho\|_F &\le&  \|X_0 - \rho\|_F \gamma_0 \gamma_1 
                + \omega \mu_2,   \\
        \|X_3 - \rho\|_F &\le&  \|X_0 - \rho\|_F 
            \gamma_0 \gamma_1 \gamma_2 + \omega \mu_3,   \\
            &\vdots&    \\
        \|X_k - \rho\|_F &\le&  \|X_0 - \rho\|_F 
          \gamma_0 \gamma_1\cdots\gamma_{k-1} 
            + \omega \mu_k. 
    \end{array}
    \label{Eq:xk_recursion}
  \end{equation}
  Note the RIP satisfies $\delta_r \le \delta_{r'}$ for $r\le r'$
  so that $\delta_{2r}\le\delta_{3r}$ and $\frac{1}{1-\delta_{2r}} \le \frac{1}{1-\delta_{3r}}$.
  We now show the relation by induction.
  \begin{itemize}
      \item For $k=1$ step, we used the initialization Eq.\ (\ref{Eq:initX0}) and recursion Eq.\ (\ref{Eq:NewStepBd}) to verify
      \[  \|X_1 - \rho\|_F \le 
        \left(\theta + \frac{4\delta_r}{\sigma_r}\|\rho\|_F   
      + 4\sqrt{2r} \frac{\lambda}{\sigma_r} \right) \|X_0 - \rho\|_F 
      +  \frac{2\sqrt{2r}\lambda}{1-\delta_{2r}} 
         \le (\theta + \eta + \phi ) \|X_0 - \rho\|_F  + \omega,
      \]
      as desired, where we have used $\|\rho\|_F\le \sqrt{r}\sigma_1$ since $\rho$ is of rank $r$.
      
      \item For $k=2$ step, we still use Eq.\ (\ref{Eq:NewStepBd}) and the definition of $\gamma_1$ and $\mu_2$ to show
        \[  \begin{array}{ccl}
            \|X_2 - \rho\|_F &\le& \left(\theta + \frac{2}{\sigma_r}
                \left( \|X_0 - \rho\|_F \gamma_0 + \omega \right)
                \right) \| X_1 - \rho\|_F + \omega \\
            &\le& \left(\theta + \frac{2}{\sigma_r}
                \left(2\delta_{2r}\|\rho\|_F + 2\sqrt{2r}\lambda \right)  \gamma_0 
                + \frac{\phi}{1-\delta_{2r}} \right)
               \| X_1 - \rho \|_F + \omega \\ 
            &\le& \left(\theta + (\eta + \phi)  \gamma_0 
                + \frac{\phi}{1-\delta_{2r}} \right)
              \| X_1 - \rho \|_F + \omega \\ 
            &\le& \gamma_1 \left(\gamma_0 \| X_0 - \rho \|_F + 
                    \omega \right) +  \omega
             = \| X_0 - \rho \|_F \gamma_0 \gamma_1 + \omega(1+\gamma_1),
        \end{array}
      \]
    as desired, where in the last inequality we have used the result of $k=1$. The recursions 
      \[  \gamma_1 = \theta + (\eta + \phi)  \gamma_0 
                + \frac{\phi}{1-\delta_{2r}}  \text{ and }
          \mu_2 = 1+ \gamma_1
      \] 
     are as defined and thus verified.
      
      \item Assume the $k$-th step satisfies the relation
        \[  \|X_k - \rho\|_F \le  \|X_0 - \rho\|_F 
                        \gamma_0 \gamma_1\cdots\gamma_{k-1} 
                + \omega \mu_k,   
        \]      
        and the recursion relation for $\gamma_k$ obeys
        \[  \gamma_k = \theta + (\eta+\phi)\times
                  \gamma_0\gamma_1\cdots\gamma_{k-1}
                    + \frac{\phi}{1-\delta_{2r}}\times \mu_k.
        \]
      \item Then according to Eq.\ (\ref{Eq:NewStepBd}), the $(k+1)$-th step will be 
      \begin{equation}
        \begin{array}{ccl}
            \|X_{k+1} - \rho\|_F &\le& \left(\theta + \frac{2}{\sigma_r}
                \left( \|X_0 - \rho\|_F  \gamma_0 \gamma_1\cdots\gamma_{k-1} 
                + \omega \mu_k \right) \right)
               \| X_k - \rho \|_F + \omega \\
             &\le& \left(\theta + \frac{2}{\sigma_r}
                \left( 2 \delta_{2r}\sqrt{r}\sigma_1 
                        + 2\sqrt{2r}\lambda \right) 
                    \gamma_0 \gamma_1\cdots\gamma_{k-1} 
                + \frac{\phi}{1-\delta_{2r}}\mu_k \right)
               \| X_k - \rho\|_F + \omega \\
             &\le& \left(\theta + 
                (\eta + \phi) \gamma_0 \gamma_1\cdots\gamma_{k-1} 
                + \frac{\phi}{1-\delta_{2r}}\mu_k \right)
               \| X_k - \rho\|_F + \omega \\ 
             &=& \gamma_k \| X_k - \rho\|_F + \omega   \\
             &\le& \gamma_k \left( \|X_0 - \rho\|_F 
                        \gamma_0 \gamma_1\cdots\gamma_{k-1} 
                + \omega \mu_k   \right) + \omega   \\
             &=&   \|X_0 - \rho\|_F 
                        \gamma_0 \gamma_1\cdots\gamma_k 
                + \omega ( 1+ \mu_k \gamma_k), 
        \end{array}
        \label{Eq:x(k+1)induction}               
      \end{equation} 

      as desired, where the first equality uses the $\gamma_k$ recursion in the $k$-th step.
  \end{itemize}
  We complete the relation of these bound in terms of $\|X_0 - \rho\|_F$. 
  
  Finally, we show the existence of sufficient conditions to minimize estimation errors $\|X_k - \rho\|_F$. In other words, we show that with small enough noise $\lambda$ and $\delta_{3r}$ we have
  the existence of the upper bound of each $\gamma_i\le \bar{\gamma} <1$ such that the bound due to $\|X_0 - \rho\|_F$ can be minimized to zero.
  
  With small noise $\lambda \le \sigma_r/(20\sqrt{2r})$ that is $\phi\le 1/5$, we show that
  \[     \delta_{3r}\le
            \frac{\sigma_r}{\sigma_1}\frac{1}{80\sqrt{r}}
  \]
   suffices for the existence of the upper bound $\bar{\gamma}<1$. 
   Since $\delta_{2r} \le \delta_{3r}$, this in turn gives $\eta\le 1/20$. Besides $\delta_{3r} \le\frac{1}{80}$ so
  \[ \theta = \frac{4\delta_{2r} + 2\delta_{3r}}{1-\delta_{2r}}
           \le \frac{6\delta_{3r}}{1-\delta_{3r}} \le \frac{6}{79}.
  \]
  In this case, we have $\gamma_0\le 0.3259, \gamma_1 \le 0.3599, \gamma_2\le 0.3807$ and $\gamma_3\le 0.3945$ and so on. 
  In the spirit of induction we can also show the existence of upper bound $\bar{\gamma}$. Suppose $\gamma_0, \gamma_1, \cdots, \gamma_k < \bar{\gamma}$ for some $\bar{\gamma} < 1$. Then we have 
  \[ \mu_k < 1 + \bar{\gamma} + \cdots \bar{\gamma}^{k-1} < \frac{1}{1-\bar{\gamma}}.
  \]
  Define
  \[  A_k := \theta + (\eta+\phi)\times
                              \gamma_0\gamma_1\cdots\gamma_{k-1}
                    + \frac{\phi}{1-\delta_{3r}}\times \frac{1}{1-\bar{\gamma}}.
  \]
  Then the condition of $ A_k < \bar{\gamma}$ for some $k = k_0$ suffices to show the existence of $\bar{\gamma}$, since in this case we have
  \[  \begin{array}{ccl}
      \gamma_{k_0 +1} &=& \theta + (\eta+\phi)\times
                              \gamma_0\gamma_1\cdots\gamma_{k_0}
                    + \frac{\phi}{1-\delta_{3r}}\times (1+\gamma_{k_0}\mu_{k_0}) \\
        &<& \theta + (\eta+\phi)\times
                        \gamma_0\gamma_1\cdots\gamma_{{k_0}-1}
                    + \frac{\phi}{1-\delta_{3r}}\times (1+\bar{\gamma} + \cdots \bar{\gamma}^{k_0}) \\
        &<& A_{k_0},            
  \end{array}
  \]
  which will automatically satisfy $\gamma_{{k_0}+1} < \bar{\gamma}$ and therefore $\gamma_k < \bar{\gamma}\ \forall k > k_0$. 
  
  In the case of $\lambda \le \sigma_r/(20\sqrt{2r})$ and 
       $\delta_{3r}\le 
            \frac{1}{80}\frac{1}{\kappa\sqrt{r}}$,
  where $\kappa = \frac{\sigma_1}{\sigma_r}$ is the condition number of the underlying density matrix $\rho$, we have
  $\gamma_0\le 0.3259, \gamma_1 \le 0.3599, \gamma_2\le 0.3807$ and $\gamma_3\le 0.3945$ all smaller than $0.45$. We can verify that $A_4 = 0.4486 < 0.45$ so that $\bar{\gamma} < 0.45$ so the existence of $\bar{\gamma}<1$ is guaranteed.
  
  It can be shown that with smaller noise bound $\lambda$, the requirement of the RIP constant $\delta_{3r}$ is less strict and it can always guarantee the existence of $\bar{\gamma} < 1$. 
  For noise $\lambda \le \sigma_r/(40\sqrt{2r})$ that is $\phi\le 1/10$, then $\delta_{3r}\le \frac{1}{20}
  \frac{1}{\kappa\sqrt{r}}$ leads to 
  $\gamma_0\le 0.6157, \gamma_1 \le 0.6057, \gamma_2\le 0.5967$,  $\gamma_3\le 0.5887$, $\gamma_4\le 0.5817$ and $\gamma_5\le 0.5757$ all smaller than $0.62$. With $A_5 = 0.6156 < 0.62$, we have $\bar{\gamma} < 0.62$ for this case.
  
  The condition $\delta_{3r} \le \frac{C}{\kappa\sqrt{r}}$ translates into the Pauli sampling requirement $m \ge C_2 \kappa^2 r^2 d \log^6 d$ to make the mapping $\mathcal{A}$ has RIP with high probability, according to Theorem \ref{Thm: RIP_pauli}. This completes the proof. 
\end{proof}

\section{Numerical Experiment\label{S:numerical}}
%\mh{Require expansion. One paragraph for the environment. One paragraph to elaborate the figure. One paragraph to compare ours with known results.}

%{\color{blue} Reference 27 = 33}

To implement our algorithm, we first need to get the $y  =\mathcal{A}(\rho)$ which is the Pauli expectation value. However, we can only collect the outcomes of Pauli basis measurement. We use the open-source software Qiskit \cite{Qiskit} and IBM quantum simulator by measuring the quantum state on a Pauli basis and recording the outcomes. For each Pauli operator, we take $l=$8192 shots and use outcomes to compute the expectation value of such Pauli as our initial $y.$ 
To connect the measurement outcomes to the expectation value of the Pauli operator $S$, we simple use
\begin{equation}
\operatorname{Tr}(S\rho) \approx \sum_{l\in\{0,1\}^{k}}(-1)^{\chi(l)}
\end{equation}
where $k$ is the qubit number, $d=2^k$ and $l$ is our measurement output (each output is a $0,1$ bit string with length $k$, $\ket{1}$ eigenstate corresponds to eigenvalue $1$ and $\ket{0}$ eigenstate corresponds to eigenvalue $-1$ for Pauli $\sigma_1,\sigma_2,\sigma_3$. For identity $\sigma_0$, both corresponds to eigenvalue $1$). So we define $\chi(l):\{0,1 \}^k \to \mathbb{N} \cup \{0\}$:  \begin{equation}
    \chi(l)=\sum_{i=1}^{k} \chi_{\scaleto{S_i}{5pt}}(l_i)
\end{equation}
where $\chi_{\sigma_0} =0$ and $\chi_{\sigma_1}(l_i)=\chi_{\sigma_2}(l_i)=\chi_{\sigma_3}(l_i)=l_i.$ For example, suppose $S=\sigma_0\sigma_1\sigma_2\sigma_3,$ and $l=\{1100\},$ we have $\chi(l)=\sum_{i=1}^{4} \chi_{\scaleto{S_i}{5pt}}(l_i)=1.$

After collecting data and estimating the $y$, we then conduct our RGD algorithm \ref{alg:RGD} and compare it with the latest non-convex optimization method called Momentum-Inspired Factored Gradient Descent (MIFGD) \cite{kim2021fast}. To the best of our knowledge, MIFGD is the best non-convex algorithm solving Eq.\ \eqref{Eq:A} till now shown in the quantum tomography literature. As they show \cite{kim2021fast}, non-convex method performs better than convex method according to their results. 
MIFGD is the updated version of the so-called Projected Factored Gradient Decent  (ProjFGD) \cite{kyrillidis2018provable}. ProjFGD performs gradient descent over the $A$ variable by writing $\rho=AA^{\dagger}$ for $A\in \mathbb{C}^{d\times r}$ and performing the following optimization:
\begin{equation}
    \displaystyle \min_{A\in\mathbb{C}^{d \times r}}\
        f(X) := \frac{1}{2}\|y-\mathcal{A}(AA^{\dagger})\|_2^2\ \ \text{subject to } 
        \| A \| _{F}^2 \leq 1
    \label{Eq.OptProb}
\end{equation}
where $\mathcal{A}$ is the same as Eq.\ \eqref{Eq:A}.
The updated MIFGD uses a refined version of the update rule \cite{kim2021fast} using Factored Gradient Descent (FGD) algorithm \cite{park2018finding_NCMF_provably}. We refer to the detailed updated rule in the original paper \cite{kim2021fast}.

We consider both the Hadamard state: $\text{Hadamard}(k)=(\frac{\ket{0}+\ket{1}}{2})^k$, and the GHZ state: $\text{GHZ}(k)=\frac{\ket{0}^n+\ket{1}^k}{2^k}$, where $k=6,8$, as our ground true state to recover. In order to compare with MIFGD, we pick the following number of Pauli measurements similar to them. For Hadamard$(6)$ state, we use $m=819 \approx 0.2\times 4^6 $ Pauli measurements and for GHZ$(6)$ state, we use $m=1638 \approx 0.4\times 4^6$ Pauli measurements. For Hadamard$(8)$ state, we use $m=13107 \approx 0.2\times 4^8$ Pauli measurements. For GHZ$(8)$ state, we use $m=26214\approx0.4\times 4^8$ Pauli measurements. For all four experiments, we use total $l=8192$ shots for each measurements. We pick hyperparameters momentum $\mu \in \{1/8, 1/4,1/3,1/2,3/4 \}$ and step size $\eta=0.01$ as \cite{kim2021fast}. We refer the definition of hyperparameters to \cite{kim2021fast}.

\begin{figure}
	\centering
	\includegraphics[width=\textwidth]{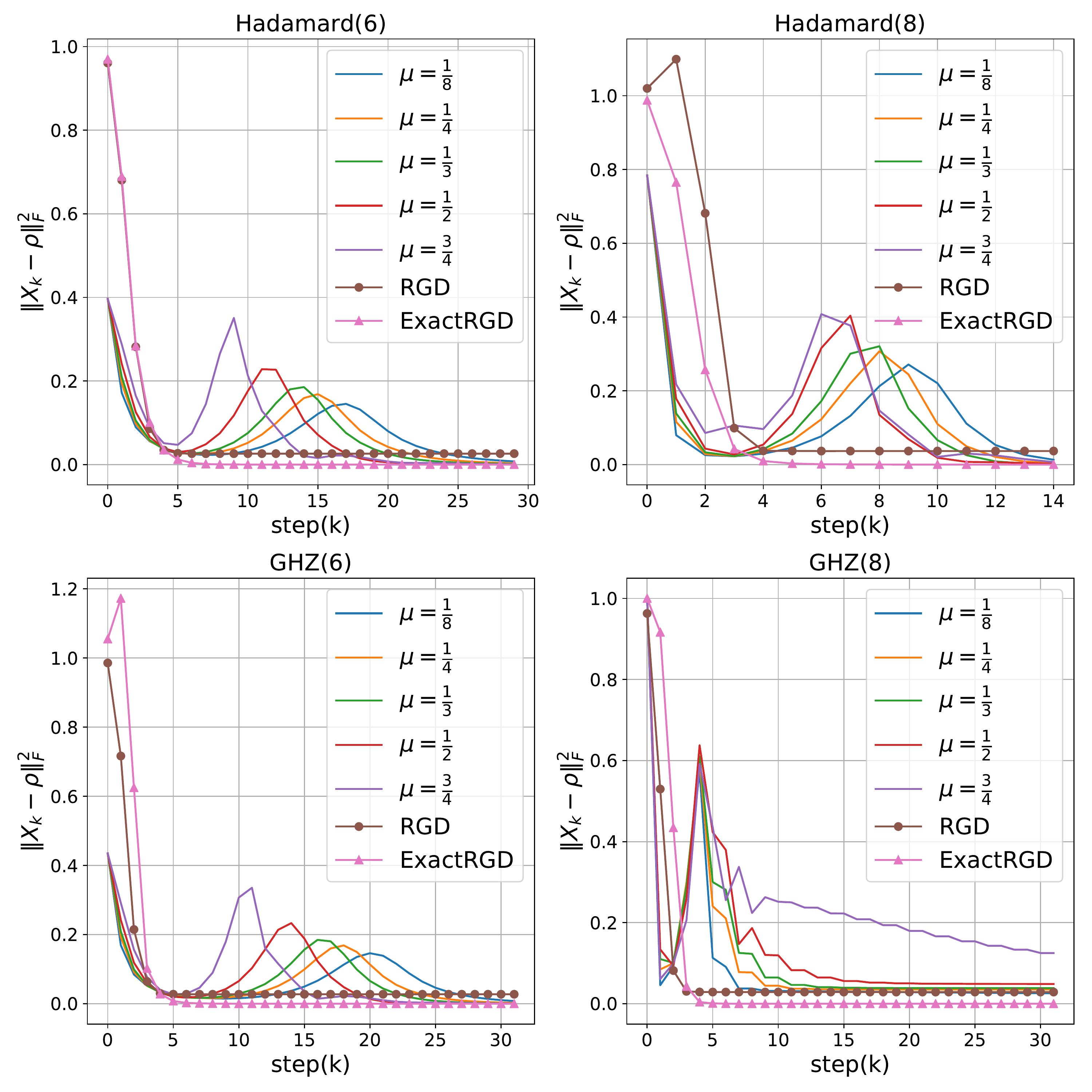}
	\caption{Target error list plots $\| X_k-\rho \|^2_{F}$, where $\rho$ is the true density matrix. For Hadamard$(6)$ state, we use $m=819\approx 0.2\times 4^6$ Pauli measurements and $l=8192$ shots. For GHZ$(6)$ state, we use $m=1638\approx 0.4\times 4^6$ Pauli measurements and $l=8192$ shots. For Hadamard$(8)$ state, we use $m=13107\approx 0.2\times 4^8$ Pauli measurements and $l=8192$ shots. For GHz$(8)$ state, we use $m=26214\approx0.4\times 4^8$ Pauli measurements and $l=8192$ shots. We choose all  $\eta=0.01$. We use the dots on our plots for both RGD and ExactRGD. Our converging speed is much faster than the MIFGD method for all $\mu.$}
	\label{fig:tomo}
\end{figure}

Our result is summarized in Fig.~\ref{fig:tomo}. The $x$ axis is the time step and the $y$ axis is approximation error between the reconstructed matrix and the true density matrix. 
We also perform the RGD algorithm on the exact data (ExactRGD) $y$ which comes from the direct calculation of the expected value of Pauli operators. We can see that the exact calculation converges super fast and the error indeed goes to $0.$

We can see our converging speed is much faster than the MIFGD method for all $\mu$ and also much more stable. The final error converges to the range of $[0.01,0.03]$ comparable to their results. The error converges at the same order.

\section{Conclusion\label{S:conclusion}}
The density matrix of interest in quantum systems is most likely of low rank, and therefore the tomography in recovering the density matrix can be formulated as a non-convex problem.
Under the compressed sensing framework, we solved the quantum tomography problem using the Riemannian gradient descent (RGD) approach. 
It is very efficient using RGD to solve the non-convex tomography formulation, since the algorithm utilizes the low rank structure directly and searches for the solution over the tangent space of low rank. 

The estimated matrix via RGD is updated iteratively and in each step the error is minimized with a multiplicative contracting factor. This gives an exponential decrease of the errors. Most importantly, the contracting factor is universal and independent of the condition number, the rank, and so on. Therefore, the number of convergent steps is logarithmic with respect to the final error.
In the noiseless case, the estimated density matrix can be arbitrarily close to the true underlying density matrix, since
the their Frobenius norm difference can be reduced nearly to zero with arbitrary precision.
For the noisy recovery of matrices, we theoretically proved that in the small noise regime, our RGD approach can converge to the true matrix density with nearly optimal bound difference. 
%Near optimal bound means that both in the Frobenius norm and the nuclear norm the errors are at the same order as those bounds theoretically obtained from convex optimization approach.
Moreover, we quantify regimes of small noise, RIP constant, and number of measurements required such that the conditions of the RGD algorithm are required to achieve this theoretical bound.
Numerical results show the largely improved performance in efficiency  for our RGD approach when compared to other approaches, including the other non-convex approaches.

% =================================================================== %

%\printbibliography%<-print bibliography

%bibliographystyle{abbrv}
\bibliographystyle{unsrt}
%\bibliographystyle{chicago}
%\spacingset{1}
\bibliography{IISE-Trans}

\appendix

\section{Appendix}

\subsection{The proof of Lemma \ref{Lem:Az_bound} \label{Appendix:z_noise}}
\begin{proof}
  This proof basically follows from \cite{Flammia_2012, Tropp_2020fastLS} with some modification.

Once each $S_i\in\{W_1, \cdots, W_{d^2}\}$ is sampled, we get the approximated coefficients $\operatorname{Tr}(S_i\rho)$ from 2-outcome measurement $\{\frac{I+S_i}{2}, \frac{I-S_i}{2}\}$. The measurement outcome is a random variable $Z_i$ with subscript index $i$ corresponding to the Pauli matrix $S_i$. 
Each instance of the random variable $Z_i$ (from the 2-outcome measurement result) is denoted as $Z_i^j (i\in[m], j\in[l])$ with the superscript $j$ refers to the $j$-th instance and we do $l$ times of measurements for each $S_i$.
The instance $Z_i^j = +1$ occurs with probability $\operatorname{Tr}(\frac{I+S_i}{2}\rho)$, while $Z_i^j= -1$ occurs with probability $\operatorname{Tr}(\frac{I-S_i}{2}\rho)$. The frequency average denoted $f_i$ is $f_i = \frac{1}{l}\sum_{j=1}^{l}Z_i^j$, and its  expectation value is
$\mathbb{E}[f_i] = \mathbb{E}[\mathbb{Z}_i] = \operatorname{Tr}(\frac{I+S_i}{2}\rho) - \operatorname{Tr}(\frac{I-S_i}{2}\rho) = \operatorname{Tr}(S_i\rho)$. 

For the input vector $y\in\mathbb{R}^m$ needed for the map $\mathcal{A}^\dagger$, we associate each component the scaled frequency average defined by 
$ y_i := \sqrt{\frac{d}{m}}f_i$. 
Therefore, according to $\mathbb{E}[f_i]$ and the operator $\mathcal{A}$ defined in Eq.\ (\ref{Eq:A}),  we know that 
$\mathbb{E}(y) = \mathcal{A}(\rho)$. We can then write $y = \mathcal{A}(\rho) + z$, where $z\in\mathbb{R}^m$ represents the noise due to measurements. 
Collecting all $S_i$ measurement outcomes, then according to Eq.\ (\ref{Eq:A}) and (\ref{Eq:Adagger}) we have
\[ \mathcal{A}^\dagger(y) 
    = \frac{d}{m} 
        \sum_{i=1}^m \left(\frac{1}{l} \sum_{j=1}^{l} Z_i^j \right) S_i,
\]
and 
\[ \mathcal{A}^\dagger \mathcal{A}(\rho) = \frac{d}{m} \sum_{i=1}^m \operatorname{Tr}(S_i\rho) S_i.
\]
The two equations then lead to
\[ \mathcal{A}^\dagger (y - \mathcal{A}(\rho)) = \frac{d}{ml} \sum_{i=1}^m
    \sum_{j=1}^{l} (Z_i^j - \operatorname{Tr}(S_i\rho)) S_i. 
\]
Since each $S_i$ is iid uniformly sampled from $\{W_1, \cdots, W_{d^2}\}$) and each $Z_i^j$ is an instance of random variable $Z_i$ from the 2-outcome measurements, 
we write
$\mathcal{A}^\dagger (y - \mathcal{A}(\rho)) = \sum_{i=1}^m \sum_{j=1}^{l} M_{ij}
$
as a sum of matrix-valued random variables 
\[  M_{ij} = \frac{d}{ml}(Z_i^j - \operatorname{Tr}(S_i\rho)) S_i.
\]
This allows us to apply the concentration technique (the matrix Bernstein inequalities) \cite{tropp2012user}. 
\begin{itemize}
    \item First $M_{ij}$ is verified to have zero mean, that is 
$\mathbb{E}[M_{ij}] = 0$, due to $\mathbb{E}[Z_i^j] =  \operatorname{Tr}(S_i\rho)$.

    \item Also we know the bound $\|M_{ij}\|= 2\frac{d}{ml} =: R$, since both $Z_i^j$ and $\operatorname{Tr}(S_i\rho)$ are in the range of $[-1, 1]$.
Then we have to bound the sum of the variances. Since $(S_i)^2 = \mathbf{I}$ for all $i\in[m]$, where $\mathbf{I}$ is the matrix identity, we know that
\[ \mathbb{E}[M_{ij}^2] = \left(\frac{d}{ml}\right)^2 \mathbb{E}[(Z_{i}^{j} - \operatorname{Tr}(S_i\rho))^2]\mathbf{I} = \left(\frac{d}{ml}\right)^2 [1 - \operatorname{Tr}(S_i\rho)^2]\mathbf{I}.
\]
Then, the sum of variances is bounded by 
\[ \sigma^2 = \| \sum_{ij} \mathbb{E}[M_{ij}^2] \|  = \left(\frac{d}{ml}\right)^2 \sum_{ij} [1 - \operatorname{Tr}(S_i\rho)^2] 
 \le \frac{d^2}{ml}.
\]

    \item With these ingredients, we know from the matrix Bernstein concentration that
\[ \operatorname{Pr}[\|\mathcal{A}^\dagger (y - \mathcal{A}(\rho))\| \ge \lambda] \le d\cdot \exp(-\frac{\lambda^2}{\sigma^2+(R\lambda/3)}) 
         \le d\cdot \exp(-\frac{ml \lambda^2}{d(d+1)}),  
\]
where the last inequality is from 
the assumption that 
$\lambda$ representing the error of interests is less than 1.
\end{itemize}

With $z = y - \mathcal{A}(\rho)$ and 
 $ml=C d(d+1)\log d/\lambda^2$, then 
$\|\mathcal{A}^\dagger (z)\| \ge \lambda$ happens
with probability at most $d^{1-C}$, as claimed.
\end{proof} 
 
% --------------------------------------------- %
\subsection{Lemma \ref{Lem:X0_bound}: the bound of the initial error $\|X_0 - \rho\|_F$ \label{Appendix:X0_bound}}
In Lemma \ref{Lem:X0_bound}, we show the bound of the initial error $\| X_0 - \rho \|_F$ of the chosen initial $X_0=\mathcal{H}_r(\mathcal{A}(y))$ from the input measurement data $y$ corresponding to the underlying density matrix $\rho$ of rank $r$.
This is used in Theorem \ref{Thm:rnk_r}
to show the iterate errors after iterations starting from this initial $X_0$.

\begin{lemma}
    \label{Lem:X0_bound}
    Let $y=\mathcal{A}(\rho)+z\in\mathbb{R}^m$ be the measurement result of the density matrix $\rho$ of rank $r$ under the sensing mapping $\mathcal{A}$ defined in Eq.\ (\ref{Eq:A}) and $z$ is the noise obeying $\|\mathcal{A}^\dagger(z)\| \le \lambda$. Suppose $m=O(\frac{1}{\delta^2} rd\log^6 d)$ such that $\mathcal{A}$ satisfy RIP with high probability according to Theorem \ref{Thm: RIP_pauli}.
    Let $\mathcal{H}_r$ denote the hard thresholding operator keeping rank $r$. 
    Then the choice of $X_0 = \mathcal{H}_r(\mathcal{A}(y))$ satisfies
  \begin{equation}
      \|X_0 - \rho\|_F \le 2\delta_{2r}\|\rho\|_F + 2\sqrt{2r}\lambda
      \label{Eq:initX0}
  \end{equation} 
\end{lemma}

\begin{proof}
  Here we want to bound the Frobenius norm of the difference $\|X_0 - \rho\|_F$ with the choice of 
  $X_0 = \mathcal{H}_r(\mathcal{A}^\dagger(y))$.
  
  Since we are interested in distance between matrices $X_0$ and $\rho$ both of rank $r$, we define the spanning of their column spaces as $Q_0\in\mathbb{C}^{n\times 2r}$ and the corresponding projection $\mathcal{P}_{Q_0}$ such that $\mathcal{P}_{Q_0}(X_0) = X_0$ and $\mathcal{P}_{Q_0}(\rho) = \rho$. Therefore we can decompose matrices into 
  components in $Q_0$ and the orthogonal complement $Q_0^\perp$, resulting in 
  \[  \begin{array}{ccl}
  \|X_0 - \mathcal{A}^\dagger(y)\|_F^2 
  &=& \|X_0 - \mathcal{P}_{Q_0}\mathcal{A}^\dagger(y)\|_F^2 
  + \|(\mathcal{I}- \mathcal{P}_{Q_0}) \mathcal{A}^\dagger(y)\|_F^2, \text{ and } \\
  \|\rho - \mathcal{A}^\dagger(y)\|_F^2 
  &=& \| \rho - \mathcal{P}_{Q_0}\mathcal{A}^\dagger(y)\|_F^2 
  + \|(\mathcal{I}- \mathcal{P}_{Q_0}) \mathcal{A}^\dagger(y)\|_F^2. 
  \end{array}
 \] 
 Besides, since $X_0 = \mathcal{H}_r(\mathcal{A}^\dagger(y))$, the  Eckart-Young theorem gives $\|X_0 - \mathcal{A}^\dagger(y)\|_F^2 \le \| \rho - \mathcal{A}^\dagger(y)\|_F^2$ and therefore we know that
  \begin{equation}
    \| X_0 - \mathcal{P}_{Q_0} 
            \mathcal{A}^\dagger(y)\|_F \le
      \|\rho - \mathcal{P}_{Q_0}
            \mathcal{A}^\dagger(y)\|_F.
    \label{Eq:x0_EckartThm}
  \end{equation}
  
  From triangular inequality, it follows that
  \begin{equation}
    \begin{array}{ccl}
        \|X_0 - \rho\|_F &\le&  
     \| X_0 - \mathcal{P}_{Q_0}\mathcal{A}^\dagger(y) \|_F +
     \|\rho - \mathcal{P}_{Q_0}\mathcal{A}^\dagger(y) \|_F
      \le 2\|\rho - \mathcal{P}_{Q_0}\mathcal{A}^\dagger(y)\|_F \\
     &=&  2\| \mathcal{P}_{Q_0}(\rho) -       
     \mathcal{P}_{Q_0}\mathcal{A}^\dagger\mathcal{A}(\rho) - \mathcal{P}_{Q_0}\mathcal{A}^\dagger(z) \|_F     \\
     &\le& 2 \|(\mathcal{P}_{Q_0} -    
      \mathcal{P}_{Q_0}\mathcal{A}^\dagger\mathcal{A}
      \mathcal{P}_{Q_0})(\rho) \|_F + 2 \|\mathcal{P}_{Q_0}\mathcal{A}^\dagger(z) \|_F. 
    \end{array}
    \label{Eq:x0_triangular}
  \end{equation} 

  The first term is from noiseless term and is bounded from
  \begin{equation}
  \begin{array}{ccl}
        \|(P_{Q_0} - P_{Q_0}\mathcal{A}^\dagger
            \mathcal{A}\mathcal{P}_{Q_0})\|_F
    &=& \displaystyle 
        \sup_{\|M\|_F = 1} |\langle(P_{Q_0} -     
        P_{Q_0}\mathcal{A}^\dagger\mathcal{A}
        \mathcal{P}_{Q_0})(M), M\rangle |  \\
    &=& \displaystyle
        \sup_{\|M\|_F = 1} | \|P_{Q_0}(M)\|_F^2 - 
       \|\mathcal{A}P_{Q_0}(M)\|_F^2  | \\
    &\le& \displaystyle
        \sup_{\|M\|_F = 1}  \delta_{2r}
    \|P_{Q_0}(M)\|_F^2  \le \delta_{2r},
  \end{array}
  \label{Eq:x0_noiseless}
  \end{equation}
  where the last two inequality follows from $\mathcal{A}$ having RIP applied to the space $Q_0$ of rank at most $2r$. 
  
  For the second term due to noise $z$, we also use the fact that 
  $\operatorname{rank}(\mathcal{P}_{Q_0}(M))\le 2r$ for any matrix $M\in \mathbb{C}^{d\times d}$ 
  so that 
  \begin{equation}
    \|\mathcal{P}_{Q_0}\mathcal{A}^\dagger(z)\|_F
    \le \sqrt{2r}\|\mathcal{P}_{Q_0}
            \mathcal{A}^\dagger(z)\| 
    \le \sqrt{2r} \|\mathcal{A}^\dagger(z)\| 
    \le \sqrt{2r} \lambda,
    \label{Eq:Q0(z)bound}      
  \end{equation}
  where the last inequality comes the assumption on noise bound. Collecting the two bounds into Eq.\ (\ref{Eq:x0_triangular}), we have 
  \[ \|X_0 - \rho\|_F \le 2\delta_{2r}\|\rho\|_F + 2\sqrt{2r}\lambda
  \]
  as claimed.
\end{proof}

\end{document}